\pdfoutput=1

\documentclass[preprint,12pt,authoryear]{elsarticle}

\newif\iftwocol
\twocolfalse

\usepackage{amsmath,amssymb}
\usepackage{graphicx}
\usepackage{amsmath}
\usepackage{amssymb}
\usepackage{booktabs}
\usepackage{array}
\usepackage{tabularx}
\usepackage[T1]{fontenc}
\usepackage{lineno}
\usepackage{xcolor}
\usepackage{textcomp}
\usepackage{gensymb}
\usepackage{url}
\usepackage{hyperref}
\usepackage{xurl}
\usepackage{tikz}
\usetikzlibrary{positioning,arrows.meta,shapes.geometric,shapes.misc,calc,fit,backgrounds,decorations.pathreplacing,patterns}

\definecolor{distalcol}{HTML}{2C6FBB} 
\definecolor{somacol}{HTML}{C0392B}   
\definecolor{proxcol}{HTML}{1E8449}   
\definecolor{accent}{HTML}{B7791F}    
\definecolor{ink}{HTML}{1F2933}       
\definecolor{muted}{HTML}{5B6B7B}     

\tikzset{
  >={Stealth[length=2.4mm]},
  every node/.style={font=\small},
  dombox/.style 2 args={rounded corners=2pt, draw=#1, line width=0.9pt,
      fill=#1!7, text=ink, inner sep=5pt, align=center, minimum height=#2},
  notebox/.style={rounded corners=2pt, draw=muted!60, line width=0.5pt,
      fill=muted!5, text=ink, inner sep=4pt, align=left, font=\footnotesize},
  procbox/.style={rounded corners=2pt, draw=ink!55, line width=0.8pt,
      fill=white, text=ink, inner sep=5pt, align=center},
  flow/.style={-{Stealth[length=2.2mm]}, line width=0.9pt, draw=ink},
  flowthin/.style={-{Stealth[length=1.8mm]}, line width=0.6pt, draw=muted},
  caclabel/.style={font=\footnotesize\itshape, text=muted},
}



\journal{Neural Networks}

\newcolumntype{L}[1]{>{\raggedright\arraybackslash}p{#1}}

\newcommand{\figwidth}{\linewidth}
\iftwocol
  \newenvironment{wfig}{\begin{figure*}[tbp]\centering}{\end{figure*}}
  \newenvironment{wtab}{\begin{table*}[tbp]\centering}{\end{table*}}
  \renewcommand{\figwidth}{\textwidth}
\else
  \newenvironment{wfig}{\begin{figure}[htbp]\centering}{\end{figure}}
  
\fi

\newsavebox{\pointboxcontent}
\newenvironment{pointbox}[1]{%
  \par\medskip\noindent
  \begin{lrbox}{\pointboxcontent}%
  \begin{minipage}{\dimexpr\textwidth-2\fboxsep-2\fboxrule\relax}%
  \medskip
  {\bfseries #1}\par\smallskip
}{%
  \medskip
  \end{minipage}%
  \end{lrbox}%
  \noindent\fbox{\usebox{\pointboxcontent}}%
  \par\medskip
}

\newif\ifarchivebackmatter \archivebackmattertrue

\input{preamble-preprint}

\begin{document}

\begin{frontmatter}

\title{From anatomy to computation: axonal delay dispersion shifts neurons from event to sequence detection}

\author[inst1]{Cheng Bi}
\ead{bichengbj@ustc.edu}
\author[inst2]{Jipeng Sun\corref{cor1}}
\ead{jipeng.sun@princeton.edu}
\cortext[cor1]{Corresponding author}

\affiliation[inst1]{organization={Independent Researcher}}
\affiliation[inst2]{organization={Department of Computer Science, Princeton University}}

\begin{abstract}
Cortical neurons fire sparsely---often fewer than one spike per sensory
window---so within that window a neuron's spike count carries little
information, and its spike timing can carry more.
That conduction delays convert firing order into synchrony is long established,
from delay-line models to polychronization.
What governs \emph{which class} of temporal feature a given neuron
detects---one volley of coincident input, or two in one particular order---has
not been examined.
We propose a delay-signature framework in which the set of axonal conduction
delays converging on a dendritic branch or a small neuron's soma constitutes a
physical key: only input sequences whose spike-time differences the delays
compensate arrive synchronously, and coincidence detection, a threshold on
input summed within an inhibition-compressed window, converts that synchrony
into an all-or-none output.
First, in simulations of an integrator-neuron model, a single physical
scalar---the dispersion of the delay set---moves a population monotonically
from event detection to order-selective sequence detection.
The transition is emergent, arising under random delays and connectivity, and
an exact calculation from the arrival statistics reproduces it: at narrow
dispersion, sequence detectors do not exist, and the dispersion at which they
overtake event detectors tracks the inter-event interval.
This maps a computational distinction onto the anatomical one between
myelinated and unmyelinated projections, making myelination a switch on what a
neuron computes.
Second, the same dispersion sets the code's limits: it bounds the longest
codable interval and fixes an absolute timing tolerance of order a
millisecond, the same at every dispersion.
Third, that millisecond window and the conduction velocity of afferents inside
cortex together set column diameter, and column anatomy predicts the window
measured in mouse barrel cortex.
One anatomically measurable parameter thus shapes what a neuron detects and
sets the limits of what it can represent.

\end{abstract}


\end{frontmatter}

\section{Introduction}

Cortical neurons fire sparsely. Ferret V1 neurons respond to natural images
at a median rate of $\sim$4.1~spikes/s, fewer than 0.5~spikes per 100~ms
perceptual window, although half of them respond to more than a quarter of the
images \citep{tolhurst2009sparseness}. In awake rat A1, a typical sound drives
fewer than 5\% of neurons to fire at high rates at any moment
\citep{hromadka2008sparse}. In mouse barrel cortex, whisker touch depolarizes
every L2/3 pyramidal cell but drives spikes in only $\sim$10\% of them
\citep{crochet2011synaptic}. Yet the tasks these neurons subserve demand
high information throughput, and at these rates a single neuron's spike count
carries little of it within one perceptual window. Lateral geniculate relay
neurons fire at 1--5~Hz spontaneously, and a typical LGN neuron emits exactly
one spike for an optimal stimulus and none for a non-optimal one
\citep{hubel1960single,reinagel1999encoding}: within one sensory window its
spike count approximates a binary variable, at most two discriminable states.
The timing of a spike adds capacity without adding spikes, which makes it the
natural carrier for a sparse code \citep{thorpe1996speed}, and sparse yet
temporally precise codes are found across modalities---delay-line sound
localization in the auditory
brainstem \citep{carr1990peripheral}, ultra-sparse sequence generation in
birdsong \citep{hahnloser2002ultrasparse}, sparse coding of natural scenes in
visual cortex \citep{vinje2000sparse}.

If information resides in the order of spikes, a downstream neuron faces a
decoding problem: how to distinguish firing orders arriving over a common set
of afferents? Heterogeneous axonal conduction delays are the classical answer.
Because axons differ in length, diameter and myelination, delays range from a
few to tens of milliseconds \citep{swadlow1985physiological}, and two spikes
emitted in one order arrive together while the reverse order arrives dispersed
(Fig.~\ref{fig:principle}a--c). This is Jeffress's delay-line model
\citep{jeffress1948place}. Hopfield \citep{hopfield1995pattern} generalized it
into the claim that delay lines converting temporal patterns into synchrony are
a substrate for pattern recognition, and the idea recurs throughout the
literature: in Abeles's synfire chains \citep{abeles1991corticonics}; in the
finding that time-dependent network properties turn temporal structure into a
spatial code \citep{buonomano1995temporal}; in Izhikevich's polychronization
\citep{izhikevich2006polychronization}; and in the demonstration that single
cortical neurons discriminate input sequences through intra-dendritic
mechanisms \citep{branco2010dendritic}.

What none of this work asks is what governs \emph{which class} of temporal
feature a given neuron becomes selective for. The distinction is concrete. One
neuron fires for a single brief volley of coincident input and ignores
everything else; call it an \emph{event detector}. Another is silent for either
of two volleys presented alone and fires only when both arrive, in one
particular order; call it a \emph{sequence detector}. Both can be built from
the same components, and existing accounts do not say which one a given neuron
becomes.

The question extends a standing one. Whether a cortical neuron is an integrator
or a coincidence detector \citep{konig1996integrator} is a question about the
width of its summation window. Which class of temporal feature it detects is a
question about the \emph{structure} inside that window. Our answer to the
second is a parameter of its own, the dispersion of the delays delivering the
input. Polychronization \citep{izhikevich2006polychronization} shows
\emph{that} sequence-selective groups form and how many can coexist; the
tempotron \citep{gutig2006tempotron}, the chronotron
\citep{florian2012chronotron} and liquid state machines \citep{maass2002real}
show that a single unit can read out spike timing through a threshold. None
identifies a physical parameter that shifts neurons from one class to the other.

We show that one scalar does---the \emph{dispersion} of the delay set converging
on a dendritic branch or a small neuron's soma, written $\Delta$ throughout---and that this scalar maps
onto the anatomical distinction between myelinated and unmyelinated
projections. A temporal code needs something that reads order, and the usual
candidate is synaptic: order shapes weights, and the weights then report how
well an input resembles what shaped them. Here the reader is the wiring itself.
Order is matched directly, by a configuration inherited from anatomy, which
makes $\Delta$ a quantity an anatomist can measure.

We call the set of conduction delays converging on one branch or soma its \emph{delay
signature}, and write it $K$: a physical configuration of anatomy that decides
which arriving patterns land together. A delay signature matches a set of
pairwise arrival \emph{differences}, so it selects an \emph{equivalence class}
of input patterns. We measure that ambiguity below: at the
delay spread attributed to unmyelinated horizontal fibres, its bound falls near
the period of the gamma rhythm.

A second consequence follows from the same scalar: a set of delays spanning
$\Delta$ cannot compensate a spike-time separation longer than $\Delta$, so
anatomy fixes the longest interval the code can represent. Selectivity and its
limit are one mechanism seen from two sides.

The closest prior work, the match-and-control principle
\citep{ellwood2024short}, also compares an arriving pattern against a stored
one within the dendrites of a single neuron and gates the output by a
threshold. Its matching substrate is short-term Hebbian potentiation, which
detects spike-train similarity; ours is a fixed physical delay configuration.
From delay learning in spiking networks
\citep{bohte2002spikeprop,shrestha2018slayer,hammouamri2024learning} and from
polychronization \citep{izhikevich2006polychronization} the framework differs
in what learning changes: there, delays are optimization variables, or form a
fixed graph over which weights are tuned; here they are biophysical givens, and
learning \emph{selects} from a redundant pool of pre-existing pathways rather
than \emph{constructing} delay values. \textbf{S5 Text} compares these and other
frameworks across five dimensions.

We formalize the resulting model---an \emph{integrator neuron} whose distal
dendrites perform delay-compensated matching, whose proximal dendrites receive
completed content, and whose soma checks one against the other---and report
three results. First, delay dispersion moves a population monotonically from
event selectivity to order-selective sequence detection, and does so with
random delays and random connectivity: the transition is emergent, and an exact
calculation from the statistics of delayed arrival reproduces it without
simulation. Second, the dispersion that buys that selectivity also
sets the limits of the code---it bounds the longest codable interval, and it
fixes the timing tolerance, of order a millisecond, that conduction must hold to.
Third, that millisecond coincidence window and the conduction velocity of
afferents inside cortex together set the diameter of a cortical column.
Read in reverse, column anatomy predicts the window, and in mouse barrel cortex,
the one area where column width, conduction velocity and window have all been
measured, the prediction matches the measurement.

Two further consequences are developed in the Discussion: the calcium threshold
is what gives the scheme a silent, tuned subpopulation at all, and the
downstream spike count that accompanies it is reported in \textbf{S3 Fig}; and
sequence coding is bounded at roughly the gamma period. Box 1 collects the
foundations the framework rests on.

\begin{pointbox}{Box 1. Framework foundations}
\noindent
Four established observations, and what each one entails:

\medskip
\noindent
Cortical neurons fire sparsely $\rightarrow$ within one perceptual window a single neuron's spike count carries little, and spike timing adds capacity without adding spikes;\\
Conduction delays are physically fixed $\rightarrow$ learning selects pathways from a redundant projection pool;\\
Myelination sets the dispersion $\Delta$ of the delays converging on a branch $\rightarrow$ $\Delta$ shapes how information is integrated downstream;\\
Coincidence detection requires a millisecond window $\rightarrow$ a calcium threshold and fast feed-forward inhibition supply it, and sparse activation is its visible signature.
\end{pointbox}
\section{Results}

\subsection{The delay-signature principle and the integrator neuron}

Consider two upstream neurons A and B projecting to a common target through
axons with conduction delays $d_A$ and $d_B$. When A fires first, its spike
travels the longer path while B's later spike travels the shorter one, and the
two arrive together; when the order is reversed the same delays add to the
firing interval, and the arrivals are dispersed
(Fig.~\ref{fig:principle}a--c). A set of delays therefore determines which
firing order is recognized and which is ignored. The principle generalizes: $N$
upstream neurons admit up to $N!$ orders, and a given delay configuration
produces synchronous convergence for only a small subset of them, as
polychronization established \citep{izhikevich2006polychronization}. What that
literature leaves open, and what the sections below tie to the dispersion of the
delay set, is which class of temporal feature a given neuron comes to detect.

The \emph{query} $Q$ is the upstream spike sequence, whose information lies in
the order in which neurons fire, and it must satisfy two conditions to be
decodable. It must be
sparse, because a dense volley always contains some pair of spikes that any
delay configuration happens to compensate, so a dense $Q$ would activate nearly
every downstream target and destroy selectivity at the population level. And it
requires a temporal reference frame for ``earlier'' and ``later'': gamma
oscillations supply one, with PV$^+$ basket cells opening a brief disinhibitory
window inside which different neurons fire at different phases
\citep{gray1989oscillatory,fries2005mechanism}. Theta phase precession in CA1
\citep{okeefe1993phase,skaggs1996theta} and the case for rank-order latency
coding along the ventral stream \citep{vanrullen2002temporal} are instances of
such phase-structured sparse firing. Sparsity preserves capacity: for $k$ of
$n$ neurons firing per sequence, order multiplies the number of
distinguishable patterns by $k!$ on top of the $\binom{n}{k}$ available from
identity alone---a factor of 6 at $k = 3$ and 120 at $k = 5$.

The \emph{key} $K_j$ is the set of conduction delays $\{d_{jf}\}$ from each
input $f$ onto branch $j$, so a spike emitted at $t_f$ arrives at
$a_{jf} = t_f + d_{jf}$. Because axons project one-to-many, a single $Q$
sequence reaches tens or hundreds of branches through as many different delay
configurations, of which only a few compensate its spike-time differences.
Learning therefore operates by \emph{selection}: in L5 distal dendrites, for example, coincident arrival triggers a dendritic calcium plateau, which induces local potentiation independently of somatic output \citep{sjostrom2006cooperative}, while chronically mismatched branches are weakened and pruned \citep{fu2012repetitive,yang2014sleep}. This
predicts sparse effective connectivity on distal branches after development,
and a dependence of what a neuron can learn
on the size of its initial redundant pool (\textbf{S1 Fig}).

\emph{Coincidence detection} converts arrival synchrony into an output. Passive
membrane time constants of 10--50~ms are far too broad for millisecond
discrimination, so the effective window must be actively compressed, and fast feed-forward inhibition is one prominent mechanism that compresses it. Where the window has been measured directly it is short: disynaptic feed-forward inhibition enforces a sub-2~ms integration window at the soma of hippocampal CA1 pyramidal cells \citep{pouille2001enforcement}, and feed-forward inhibition onto cerebellar Purkinje cells reduces the EPSP summation window to 1--2~ms \citep{mittmann2005feedforward}. It is short as well in the compartment the column prediction below concerns, and there it also moves: in slices of mouse barrel cortex, thalamocortical feed-forward inhibition holds the window over which L4 neurons integrate thalamic input to about 1~ms on a rested pathway, and repetitive thalamic stimulation widens it toward 10~ms as that inhibition depresses \citep{gabernet2005somatosensory}. Where the window has not been measured, the passive window is broad and the inhibitory circuitry that would compress it is present. In dentate granule cells, cable models fitted to dual dendritic recordings give a local summation half-width of $10.8 \pm 0.5$~ms \citep{schmidt2007single}. There, noradrenaline-enhanced CCK$^+$ feed-forward inhibition narrows the integration window from $\sim$51~ms to $\sim$4~ms \citep{glovaci2026sparsification}. L5 distal dendrites carry a high density of HCN channels \citep{berger2001high}, whose current curtails the temporal summation of distal input \citep{williams2000site}, and distal-targeting SOM$^+$ interneurons are positioned to narrow the window further. Other mechanisms, such as near-synchronous discharge among inhibitory interneurons, may also contribute to window compression in local circuits.

The motif recurs: fast inhibition, often from PV$^+$ or related subtypes, compressing the effective window to the millisecond scale where it has been measured and supplying the same circuitry where it has not. Narrow coincidence detection is therefore a general computational motif, implemented through local inhibitory microcircuits across cell types. We adopt $\tau_{\mathrm{w}} \approx 1$~ms as a canonical value throughout, the narrow end of every window measured and the L4 value on a rested pathway, and carry a 1--2~ms range through the column prediction below.

We call a neuron with three functional domains an integrator neuron
(Fig.~\ref{fig:principle}d). Its distal dendrites perform
$Q\!\cdot\!K$ matching; its proximal dendrites receive a content signal $V'$
already completed by a local attractor microcircuit
\citep{hopfield1982neural}; and its soma compares the two, in the spirit of the
index--content separation first proposed by hippocampal indexing theory
\citep{teyler1986hippocampal}. Cortical L5 pyramidal neurons are the canonical example, with distal input
from higher-order thalamus and cortex, proximal input from L2/3, and an output
reflecting whether completed content is consistent with current context;
hippocampal CA1 pyramidal neurons are a distributed implementation, with distal
matching supplied by entorhinal layer 3 and proximal index content by CA3.

Formally, branch $j$ accumulates calcium from arrivals falling within
$\tau_{\mathrm{w}}$ of a reference time $t_c$,
\begin{equation}
\mathrm{Ca}_j = \sum_{f} w_{jf}\,
\mathbf{1}_{\left\{|a_{jf} - t_c| < \tau_{\mathrm{w}}\right\}},
\label{eq:ca}
\end{equation}
and emits a hard-gated current
\begin{equation}
I_j = \eta_j\, g\!\left(\max(0, \mathrm{Ca}_j - \theta)\right),
\label{eq:hardgate}
\end{equation}
where $w_{jf}$ is a synaptic weight, $\eta_j$ the branch-to-soma transmission
efficiency, and $g$ a monotonic saturating function standing for the graded but
bounded NMDA-dependent plateau. Below $\theta$ the input contributes nothing;
above it the branch commits to a sustained current lasting 100--200~ms in
biological plateaus \citep{larkum1999new}. The gate is hard in its decision and
graded in its amplitude: once open, $I_j$ scales with the quality of the
match. The soma pools both streams,
\begin{equation}
y = \sum_{k} w_k V'_k \;+\; \alpha \sum_{j} \eta_j I_j \;-\; I_{\mathrm{inh}},
\qquad \text{fire if } y > \theta_{\mathrm{soma}},
\label{eq:soma}
\end{equation}
with $\alpha$ the global distal gain and $I_{\mathrm{inh}}$ local
inhibition. $\alpha$ is set so that a distal match strongly facilitates a
co-arriving proximal signal without being sufficient to drive firing on its
own.
The distal and somatic windows are deliberately mismatched: a $\sim$1~ms distal
window buys sequence selectivity, while the 100--200~ms plateau holds the soma
primed for the proximal signal to arrive, so that distal matching acts as a
contextual prior rather than as a direct drive
\citep{larkum1999new,larkum2004dendritic}.

Throughout this work we take axonal conduction delays as the primary source of delay dispersion; relay neurons that perform limited integration and discharge with reliable temporal precision, discussed in \textbf{S6 Text}, can serve as additional or functionally equivalent sources, because they supply a repeatable delay \emph{structure}: a stable configuration against which spike-time differences can be compensated trial after trial. Such a structure could come from axonal conduction, relay neurons, membrane integration, or any other physical process that produces stable delays, and the matching principle is the same regardless of which source supplies the delay.

\begin{wfig}
\includegraphics[width=\figwidth]{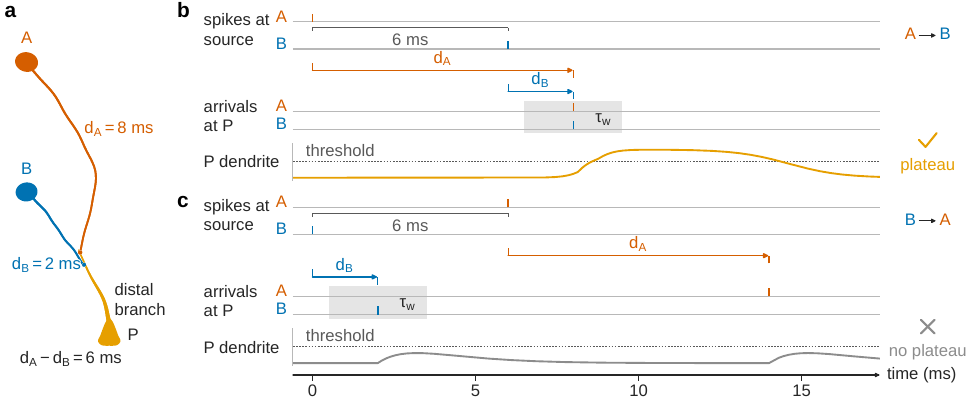}\par\vspace{1.5mm}
\includegraphics[width=\figwidth]{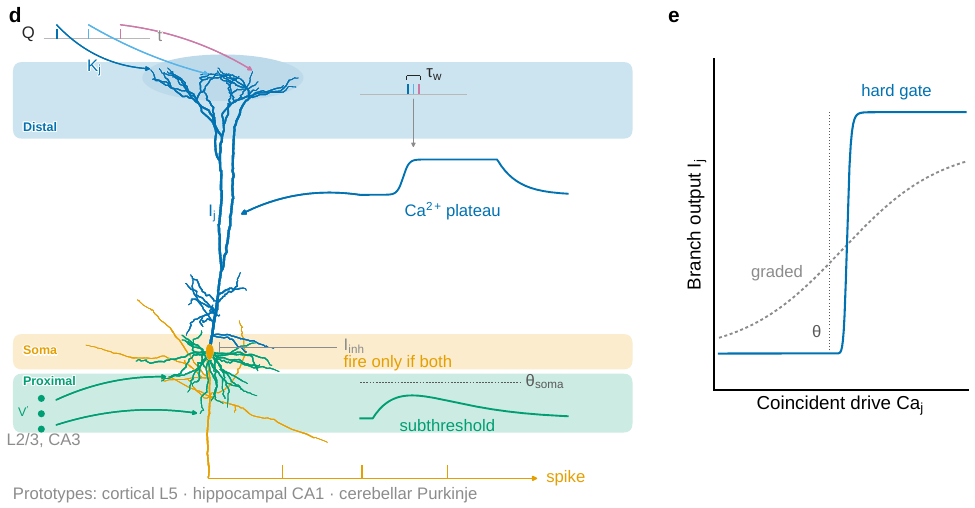}
\caption{
\textbf{The delay-signature principle and the integrator neuron.} \textbf{a}, Neurons A and B project to a common target P; A takes the longer path ($d_A = 8$~ms, $d_B = 2$~ms). \textbf{b}, A fires first, B 6~ms later. Top, firing times at the source; middle, arrivals at P, each spike displaced by its delay (bars); bottom, dendritic potential. Because $d_A - d_B$ equals the firing interval, the inputs land together inside the coincidence window $\tau_{\mathrm{w}} = 3$~ms (shaded), the summed input crosses the plateau threshold (dashed), and a plateau follows (check mark). \textbf{c}, The same delays, order reversed: they now add to the firing interval, arrivals fall 12~ms apart, and each EPSP decays subthreshold (cross mark). Panels \textbf{b} and \textbf{c} share one time axis and identical source spikes: the delays alone make the difference. \textbf{d}, A reconstructed layer-5 thick-tufted pyramidal neuron, its three functional domains shaded. The apical tuft receives long-range projections carrying the query $Q$; each branch imposes its own delay set $K_j$, and where those delays compensate the departure-time differences, spikes that left apart arrive together (rasters), the branch fires an NMDA-dependent calcium plateau (Eq.~\ref{eq:ca}), and a hard-gated vote $I_j$ travels down the bare apical trunk to the soma (Eq.~\ref{eq:hardgate}). The basal skirt receives the microcircuit-completed content signal $V'$, subthreshold alone; the soma fires only when both streams agree (Eq.~\ref{eq:soma}). \textbf{e}, The gate is hard in its decision: below $\theta$ input contributes nothing, and above it the branch commits to a sustained saturating current whose amplitude scales with the match; a soft gate passes a fraction of everything (schematic axes). Morphology: NeuroMorpho.Org (RRID:SCR\_002145), NMO\_128418, archive Kole.}
\label{fig:principle}
\end{wfig}

\subsection{Delay-signature dispersion sets detector class}

The model as described so far takes its delay configuration as given. We now
let it vary, which is where the framework's central claim lies: that the
dispersion of the
delay set---narrow for the myelinated projections of white-matter tracts, wide
for unmyelinated horizontal fibres \citep{tomassy2014distinct}---shifts a
population of neurons from event detection to sequence detection.

The direction of the effect follows from geometry.
A delay set much narrower than the interval between two events delivers each event's inputs together, so a branch that reaches threshold does so on the inputs of a single event.
A delay set as wide as the interval contains differences that cancel it: inputs of the first event that travel the long paths land with inputs of the second that travel the short ones, and a branch that neither event can bring to threshold alone fires for the pair, in that order only.
Geometry gives no more than that direction.
It does not say at what dispersion sequence detectors overtake event detectors, whether widening the delay set keeps helping, or how rare either class is when the wiring is random.
The simulations and the calculation below answer those questions, in numbers that a measured $\Delta$ can be checked against.

The link from myelination to dispersion is arithmetic. Myelination
raises conduction velocity by roughly an order of magnitude, and for a given
spread of path lengths $L$ the delays are $L/v$, so their dispersion falls as
$1/v$: faster conduction compresses the delay distribution. That cortical
pyramidal axons carry myelin in discontinuous profiles, some segments sheathed
and others bare
\citep{tomassy2014distinct}, puts both regimes inside the same cortex. If the simulations below are right about what $\Delta$ does, myelination shapes how downstream neurons integrate their input, and so which temporal features they detect.

That argument takes $v$ as uniform within a projection, so that $\Delta$ comes
from the spread of path lengths alone. Conduction velocity is itself
heterogeneous, which supplies a second source of dispersion. In mouse barrel
cortex, voltage imaging gives conduction velocities from 74 to 473~$\mu$m/ms,
depending on the layers and columns that activity travels between
\citep{scheuer2023velocity}. In monkey V1, re-analysis of earlier recordings shows that the speed grows with distance, from a median of 0.33~m/s over the short separations that dominate the sample to several metres per second beyond 5~mm \citep{zhaoping2025conduction,girard2001feedforward}. A projection therefore
does not need long axons to acquire a wide delay signature. What myelination
sets is the central tendency of $v$, and through it the scale
of $\Delta$; the residual spread of $v$ widens $\Delta$ further. Both sources
are subsumed in the simulations below, which sweep $\Delta$ directly and are
indifferent to which of the two produced it.

Populations of 1000 integrator neurons received random projections from a
200-neuron feature population, of which two disjoint sets of 40 formed two
events inside a gamma cycle, at [5,~8]~ms and [20,~23]~ms. Delays were drawn
uniformly on $[0, \Delta]$ with $\Delta$ swept continuously from 0 to 50~ms,
every population receiving identical input sequences and projection structures.
Classification used a silence check: a neuron counts as a sequence
detector only if it fires for the two events together and for neither alone.
Because the three stimuli share their spike times, the comparison is
within-neuron. A fourth stimulus---the same two volleys, same within-volley
offsets, reversed order---is presented alongside but takes no part in the
classification, so it provides an independent measure of order selectivity.

The transition is monotonic in dispersion (Fig.~\ref{fig:dispersion}). At
narrow dispersion, sequence detectors do not exist, while about one neuron in
ten responds to a single event. Sequence detectors appear as the dispersion
approaches the inter-event interval and peak beyond it, near 20~ms, before
declining as dispersion grows past the point where any pairwise difference can
be compensated. Event detectors fall monotonically over the same range, to none
at $\Delta = 30$~ms, the two classes crossing at $\Delta = 16.7$~ms. Dual
responders---neurons firing to each event alone, responsive without being
conjunctive---fall below the resolution of the sweep by $\Delta = 10$~ms, so
the neurons that wide dispersion recruits are selective. The transition and its crossing survive a time step ten times
finer, a plateau of 100~ms and spontaneous firing of the inputs, although at
the five-input gate the finer step lowers the responsive fractions two- to
threefold (\textbf{S1 Text}).

The transition follows from the statistics of delayed arrival alone, and can
be computed without simulating a neuron (Materials and methods). Whether a
neuron fires for event 1 alone, for event 2 alone or for the two together is a
question about when one dendritic potential first crosses threshold, and that
potential depends only on when the connected inputs arrive. The two events
recruit disjoint inputs whose potentials add below threshold, so the
probability of a sequence detector is the product of the probabilities of
silence to each event, less the probability of silence to both. The
calculation takes the simulation's parameters and fits none. Its curves pass
through the simulated fractions at every threshold, dispersion and interval
(Fig.~\ref{fig:dispersion}b, d), differing from them by a median of 1.8\%
wherever a class exceeds 0.3\% of the population. What the sweep measures is
therefore set by the arrival times, and so by the delays that produce them.

The reversed-order control confirms that the selectivity is for
\emph{order}. Pooled over seeds and over the
$\Delta = 14$--30~ms span in which the class exists at $\theta = 4.5$, 190 of
192 classified sequence detectors are silent to the reversed sequence (99.0\%);
over the whole 0--50~ms sweep the figure is 284 of 289 (98.3\%). The five
exceptions are scattered rather than concentrated, three of them beyond 45~ms
where the class is already collapsing. Order selectivity stays high at the
permissive gate: 92.6\% at $\Delta = 20$~ms and 89.3\% at $\Delta = 30$~ms when
$\theta = 2.5$.

Both transition points scale linearly with the inter-event interval
(Fig.~\ref{fig:dispersion}c), with slopes the theory fixes. The crossover
$\Delta^{*}$ has slope 1.16 in theory and 1.12 [1.04, 1.23] when read off the
seed-pooled simulated curves, so sequence detectors overtake event detectors
at a dispersion close to the interval, slightly beyond it at the longer
intervals. The slope exceeds one at every threshold at which the two classes
cross, and on a step ten times finer the theory's is 1.13 (\textbf{S1 Text}).
The peak $\Delta_\mathrm{opt}$ rises faster, with slope 1.33 in theory and
1.45 [1.15, 2.51] in simulation, so the dispersion that maximizes
sequence-detector yield lies beyond the interval rather than at it.

\begin{wfig}
\includegraphics[width=\figwidth]{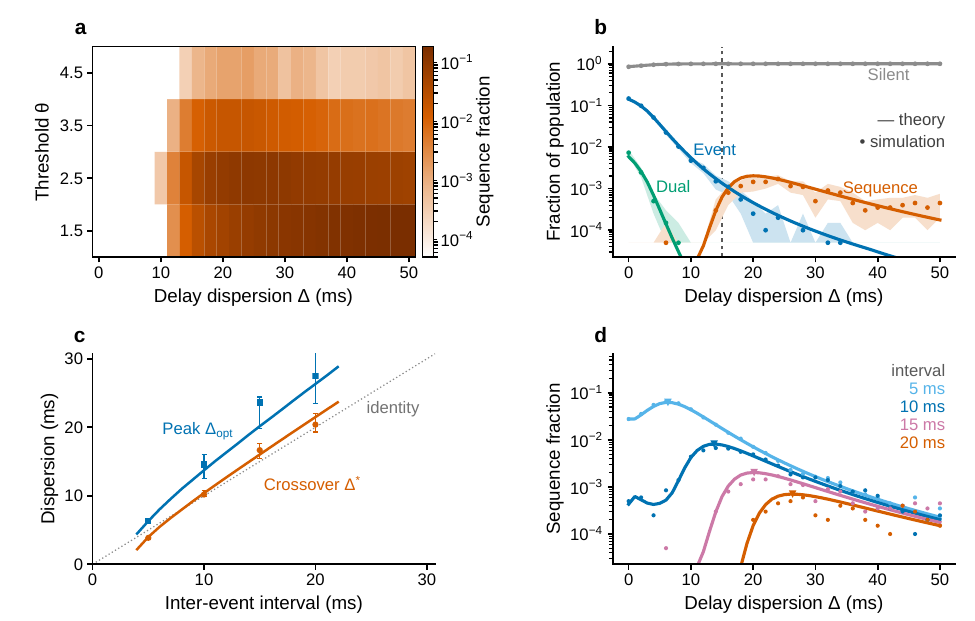}
\caption{
\textbf{Delay-signature dispersion sets detector class.} Populations of integrator neurons receive random projections from a feature population, of which two disjoint sets form two events inside a gamma cycle; conduction delays are drawn uniformly from $[0, \Delta]$. A neuron is classed as a sequence detector only if it fires for the two-event stimulus and for neither event alone. The order-reversed sequence is presented alongside but takes no part in the classification, so it provides an independent measure of order selectivity. Points are simulation, 20 seeds; lines are the analytic theory, computed from the arrival statistics with no free parameter (Materials and methods). Over all four thresholds and intervals, theory and simulation differ by a median of 1.8\% in the class fractions above 0.3\% of the population. Full parameterization in \textbf{S1 Text}. \textbf{a}, Sequence-detector fraction across dispersion and calcium threshold, at an inter-event interval of 15 ms. \textbf{b}, Detector-class fractions against dispersion ($\theta = 4.5$); shading is the 95\% bootstrap confidence interval of the simulated mean, and the dashed line marks the inter-event interval. \textbf{c}, Crossover dispersion $\Delta^{*}$, at which sequence detectors overtake event detectors, and peak dispersion $\Delta_\mathrm{opt}$, against inter-event interval, read off the seed-pooled curves with 95\% bootstrap confidence intervals over seeds. The theory gives slopes of 1.16 for $\Delta^{*}$ and 1.33 for $\Delta_\mathrm{opt}$; least-squares fits to the simulated points give 1.12 [1.04, 1.23] and 1.45 [1.15, 2.51]. \textbf{d}, Sequence-detector fraction against dispersion for each interval; triangles mark the theoretical optimum. Tuning is unimodal, and the interval sets where the optimum falls. Panels \textbf{a}, \textbf{b} and \textbf{d} use a logarithmic scale, on which the floor at $1/20{,}000$ is a single neuron in the pooled sample and marks the resolution of the sweep; simulated fractions of zero are not drawn.}
\label{fig:dispersion}
\end{wfig}

The differentiation is emergent: with random delays and random projections,
narrow-$K$ populations become uniformly event selective and wide-$K$
populations spontaneously produce sequence detectors, driven solely by a
physical parameter. The dependence has a definite direction and shape:
sparsity increases monotonically with dispersion, because
aligning two temporally separated volleys inside a millisecond window is a
stricter condition than aligning one, and strictness under random connectivity
means rarity. That yields an \emph{ordinal} account of the sparsification
gradient along the cortical hierarchy---areas whose afferents have wider delay
dispersion should host sparser, more sequence-selective responses---which is
testable against anatomy. The absolute proportions depend on connection
probability, threshold and event size (\textbf{S1 Text}); the monotonic
dependence on dispersion does not.

The decline of event detectors under wide dispersion is temporal folding: the
tail of event~1, shifted by large delays, and the head of event~2, shifted by
small ones, overlap, making cross-event coincidence more likely than
within-event coincidence. This is the ambiguity behind a delay signature's
equivalence class, the set of input patterns it brings into coincidence, seen
from the useful side. A branch cannot tell the late part of one event from the
early part of the next; when the two belong to different events, that confusion
is what builds a sequence detector, and when they belong to the same one it is
what limits the codable interval. One mechanism, two consequences; the next
section measures the second.

\subsection{The same dispersion sets the codable interval and the precision it requires}

A delay configuration that compensates one interval can compensate others, over a range set by the delays available. If the interval falls within that range, some subset of the K configurations in the redundant pool compensates it; beyond it, none does. To measure the range, we identified sequence detectors at a reference interval and probed them with the same stimulus retimed to every interval from 0 to 91~ms, holding feature identity and within-volley timing fixed so that only the separation changed.

The tuning is sharply asymmetric (Fig.~\ref{fig:limits}a). Above the reference, selectivity is excellent: the response falls to half within about a millisecond and to zero within a few. Below it the population responds broadly, forming a plateau across shorter intervals. The asymmetry follows from the redundant pool: shortening the interval creates more opportunities for some subset of the redundant delay configurations to compensate the new timing, whereas lengthening it moves the event beyond the reach of the available delays. Two statistics quantify the plateau. Averaged over probe intervals shorter than the reference, 51--54\% of detectors respond at any \emph{given} shorter interval, so below the reference the population loses its selectivity. Far fewer respond at \emph{every} shorter interval tested, 1.8--9.2\% at $\Delta = 20$~ms and 0.4--3.1\% at $\Delta = 30$~ms: different neurons match different intervals, and the code is sparse at the population level.

Both features hold across the dispersion sweep (Fig.~\ref{fig:limits}b): the plateau below the reference and the cutoff above it move together with $\Delta$.

What bounds the code is that upper cutoff, and it tracks the dispersion (Fig.~\ref{fig:limits}c). A population with delays on $[0, \Delta]$ responds to no interval longer than roughly $\Delta$: 14~ms for $\Delta = 10$~ms, 22~ms for $\Delta = 20$~ms, 31~ms for $\Delta = 30$~ms, and on to 48~ms for $\Delta = 50$~ms. Repeating the sweep at $\theta = 1.5$, 3.5 and 4.5 barely moves it: the bound is set by the delays available, which makes the upper end of the codable temporal range a property of the anatomy.

Sub-millisecond matching across a 10--30~ms delay spread also implies a precision on conduction velocity of a few per cent, and conduction velocity in vivo is temperature-, state- and activity-dependent. We identified sequence detectors with exact delays and then perturbed the delays in two ways: independent per-connection jitter, each delay scaled by its own draw of $1 + \mathcal{N}(0, s)$, modelling axon-to-axon variability; and a common scaling by $k$, modelling a global shift such as a temperature change. Because the scale factor multiplies delays, $k > 1$ is a \emph{slowing} of conduction and $k < 1$ a speeding. Perturbations this small need a finer time step than the other experiments use: on a 1~ms step a perturbation under half a millisecond rounds back to the delay it started from and costs nothing, so this sweep runs on a 0.1~ms step, where retention has converged (\textbf{S1 Text}).

The scheme degrades smoothly (Fig.~\ref{fig:limits}d, e). At $\Delta = 20$~ms, 74\% [73, 76] of sequence detectors survive 2\% velocity jitter, 57\% [56, 59] survive 5\%, and 44\% [43, 46] survive 10\%. Retention never collapses to zero even under 30\% jitter, because detectors differ in how tightly their delays happen to be matched, and the population degrades by attrition. Wider populations are uniformly more fragile, because the same fractional error on a longer delay is a larger absolute error, so wider dispersion buys sequence selectivity at a cost in precision tolerance.

Where the dispersion is close to the interval, a global scaling costs more when it speeds conduction than when it slows it. At $\Delta = 20$~ms a 5\% slowing retains 69\% [67, 70] against 62\% [60, 64] for a 5\% speeding, and a 20\% slowing 44\% [43, 46] against 22\% [21, 24]. At $\Delta = 15$~ms, equal to the interval, the gap is wider, and from $\Delta = 30$~ms on, a 5\% slowing and a 5\% speeding differ by at most five percentage points. The cause is selection. A detector is selected when its inputs arrive within the coincidence window, and where delay differences long enough to compensate the interval are scarce, the near-matches that pass mostly fall short of it: when the dispersion equals the interval, event-1 inputs arrive early in 82--84\% of sequence detectors, at each of three intervals (\textbf{S1 Text}). Dilating all delays closes that gap and compressing them widens it. Once the dispersion comfortably exceeds the interval, near-matches are as often early as late, and the asymmetry disappears.

Sweeping the dispersion from 15 to 50~ms turns the fragility of wide populations into a law (Fig.~\ref{fig:limits}f). The jitter that costs half the matched population falls as $1/\Delta$, from 7.9\% at $\Delta = 20$~ms to 2.7\% at $\Delta = 50$~ms, and repeating the measurement at 10 and 20~ms inter-event intervals moves it far less than the dispersion does. The one exception is the dispersion equal to the interval, where the detectors are the under-compensated near-matches described above and tolerate less (6.1\% at $\Delta = 15$~ms). What holds constant across the sweep is the tolerated \emph{absolute} error, 1.36~ms at $\theta = 2.5$ and 0.55~ms at $\theta = 4.5$---of the order of the coincidence window itself---while the tolerated \emph{fraction} falls as $1/\Delta$. The scheme therefore demands a fixed timing tolerance, of order a millisecond at every dispersion. Stated as a percentage of conduction velocity, the requirement is understated for wide populations and overstated for narrow ones.

The redundant pool posited by the framework is what absorbs this attrition: a population starting from hundreds of candidate branches can afford to keep only a fraction of them, and redundancy is a prerequisite for the scheme to work.

Both sweeps are reported at $\theta = 2.5$, with the five-input gate carried alongside in Fig.~\ref{fig:limits}. The strict gate narrows the equivalence class and tightens the precision budget without changing the $1/\Delta$ law; only the constant moves (\textbf{S1 Text}).

\begin{wfig}
\includegraphics[width=\figwidth]{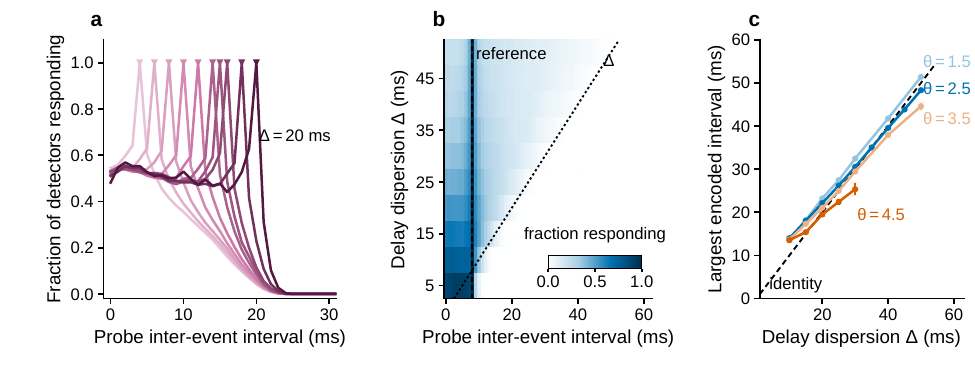}\par\vspace{1.5mm}
\includegraphics[width=\figwidth]{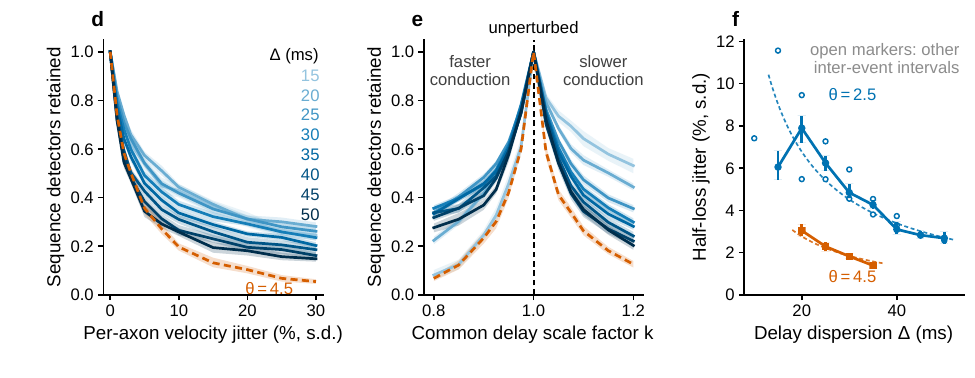}
\caption{
\textbf{The same dispersion sets the longest codable interval and the timing precision the code demands.} \textbf{a}--\textbf{c}, Sequence detectors identified at one reference inter-event interval, then probed with the same stimulus retimed to every other interval; feature identity and within-volley timing held fixed, 20 seeds. \textbf{a}, Interval tuning at $\Delta = 20$~ms, one curve per reference from 4 to 20~ms (triangles mark each curve's own reference): the response collapses within milliseconds above the reference; below it, a plateau emerges at the population level. At any given shorter interval, 51--54\% respond, but only 1.8--9.2\% respond to every shorter interval tested. Both are quoted over the 10, 15 and 20~ms references. \textbf{b}, The same measurement across dispersions at a reference of 8~ms, each row one population's tuning curve as colour: plateau and cutoff move with $\Delta$, and the cutoff edge tracks the dotted identity line. \textbf{c}, The largest interval eliciting any response ($\geq 0.01$ of detectors) against dispersion, with 95\% bootstrap confidence intervals. The bound sits just above identity---delays on $[0, \Delta]$ cannot compensate a separation longer than roughly $\Delta$---and barely moves across $\theta = 1.5$ to 4.5, so it is set by the anatomy. The strict arm stops at $\Delta = 30$~ms, where a five-input gate leaves too few detectors to measure. \textbf{d}--\textbf{f}, Precision, on a 0.1~ms step (\textbf{S1 Text}). Detectors identified with exact delays, the delays then perturbed and re-rounded to the step. \textbf{d}, Independent per-connection jitter, $1 + \mathcal{N}(0, s)$ on each delay, for dispersions of 15--50~ms (key): retention falls by attrition, and wider populations are uniformly more fragile. Dashed vermillion repeats $\Delta = 20$~ms at $\theta = 4.5$. \textbf{e}, Common scaling by $k$ ($k > 1$ lengthens delays, a \emph{slowing} of conduction): near the interval, retention is asymmetric about $k = 1$ and worse for speeding, most strongly at $\Delta = 15$~ms; from $\Delta = 30$~ms on, a 5\% slowing and speeding differ by at most five percentage points. \textbf{f}, The jitter costing half the matched population, against dispersion; filled markers the 15~ms interval of \textbf{d}--\textbf{e}, open markers 10 and 20~ms. The dashed line is the hyperbola $s_{1/2} = \bar{c}/\Delta$, not a fit. The tolerated \emph{fraction} falls as $1/\Delta$ while the tolerated absolute error holds near 1.36~ms at $\theta = 2.5$ and 0.55~ms at $\theta = 4.5$: the scheme needs a fixed timing tolerance of order a millisecond at every dispersion.}
\label{fig:limits}
\end{wfig}

\subsection{Column diameter and the coincidence window}

In the present framework, a cortical column defined by its afferents can be understood as the spatial projection of the delay-compensation mechanism: inputs converging onto the same downstream neuron are integrated in time when their conduction delay differences fall within the coincidence window---that is, when the temporal offsets of upstream spikes are compensated by differences in conduction delays---and the region on the cortical surface over which such Q$\cdot$K matching occurs defines the neuron's columnar receptive field. This spatial extent is set by the product of the afferents' conduction velocity inside cortex and the window width, $v \times \tau_{\mathrm{w}}$.

What must be compensated is the difference between two accumulated delays, so the extent of this territory depends on the angle at which the axons cross. Antiparallel axons change the arrival-time difference by $2x/v$ for a target displacement $x$, and half a window's conduction already spans the column. Parallel axons change both paths equally, and set no columnar scale at all. Orthogonal intersection lies between: only one path length changes to first order, and the column spans a full window's conduction. We adopt the orthogonal case as representative, since afferents crossing at a common target are not systematically collinear. The horizontally extending axons of L4 traverse columns rather than radiating from their centres \citep{arnold2001thalamocortical}, so $v \times \tau$ is an end-to-end span and therefore a diameter. Thalamocortical axons conduct at 0.31--0.33~m/s between the white matter and L4 \citep{salami2003change}, the only measurement of these axons inside cortex, made along their radial course in mouse slices; at that velocity,
\begin{equation}
d_{\mathrm{column}} = v_{\mathrm{axon}} \times \tau_{\mathrm{w}}
\approx 0.3~\mathrm{m/s} \times (1\text{--}2~\mathrm{ms}) = 300\text{--}600~\mu\mathrm{m}.
\label{eq:column_diameter}
\end{equation}
Fig.~\ref{fig:column}c shows the relation across conduction velocities,
and Fig.~\ref{fig:column}a the mapping from firing
latency to columnar position that the same delay gradient produces.

The afferents crossing a column need not share one speed---the dispersion
that makes the scheme work is partly a dispersion of velocities---so the single
$v$ in Eq.~\ref{eq:column_diameter} is a central tendency, and reading it that
way yields a second prediction. If $v$ varies across the afferents that
converge on a column, a column boundary set by delay compensation is graded
over a distance proportional to that variation, and columns should be more
sharply bounded in areas whose horizontal conduction is more uniform. Testing
this needs the spread of $v$ across one column's afferents, which the
measurements used here do not supply.

Horizontal conduction velocity and column width have both been measured in ten cortical areas, which \textbf{S7 Text} collects with the window each pair implies. Three of them pair a velocity that its source attributes to monosynaptic transmission with a column defined by thalamic afferents, and Fig.~\ref{fig:column}c overlays them. In macaque V1, a re-analysis of orthodromic single-unit latencies from Girard et al.\ gives local horizontal connections a median conduction velocity of 0.33~m/s \citep{zhaoping2025conduction}; against single-eye ocular dominance columns 395--670~$\mu$m wide \citep{horton1996intrinsic}, the relation implies a window of 1.2--2.0~ms. In mouse barrel cortex, the latency slope of voltage responses in parvalbumin interneurons gives an L4 intralaminar velocity of $142 \pm 76~\mu$m/ms (mean $\pm$ SD over four slices, at room temperature) \citep{scheuer2023velocity}; against barrels measured at 124--150~$\mu$m in slices \citep{judge2026biased}, a lower bound, and at about 300~$\mu$m across \citep{sofroniew2015neural}, it implies 0.9--2.1~ms. In cat area 17, field potentials evoked in vivo place horizontal conduction in the supragranular pathway at 0.3--0.6~m/s \citep{luhmann1990horizontal}; against ocular dominance columns 400--500~$\mu$m wide \citep{lowel1994ocular}, it implies 0.7--1.7~ms.

Read in reverse, the relation turns column anatomy into a prediction of the coincidence window, $\tau_{\mathrm{w}} \approx d_{\mathrm{column}}/v$, which can be checked wherever the window has also been measured. In mouse barrel cortex it has: on a rested pathway, feed-forward inhibition follows thalamic excitation onto L4 neurons by 1.2~ms and holds the window over which they integrate thalamic input to about 1~ms \citep{gabernet2005somatosensory}, inside the 0.9--2.1~ms the barrels predict. In the other areas most pairings imply 1--5~ms, a spread set largely by how velocity was measured, which moves the implied window two- to fourfold within an area (\textbf{S7 Text}). The velocity the construction needs, that of afferent arbors running horizontally in L4 over less than a millimetre, has not been measured in any area, and measuring it is what would turn this consistency into a test (Box~2).

\begin{wfig}
\includegraphics[width=\figwidth]{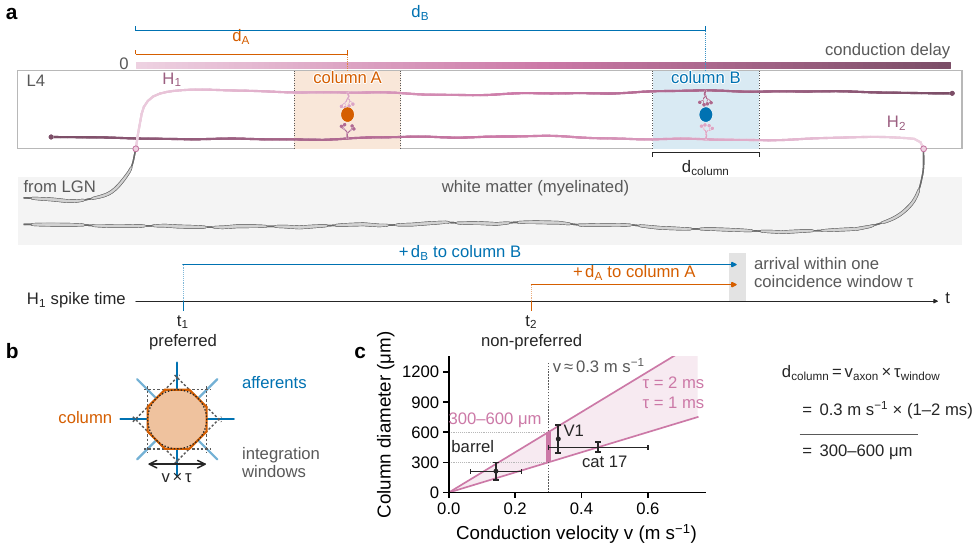}
\caption{
\textbf{The delay gradient of LGN axons maps firing latency onto columns and fixes their diameter.} \textbf{a}, Two LGN neurons with horizontally extending long axons (H$_1$, H$_2$) reach L4 through the white matter, where they are myelinated (internodal sheaths separated by nodes of Ranvier), shed that myelin at the L4 boundary, and run in opposite directions across column~A and column~B. The antiparallel arrangement maps firing latency onto columnar position; the diameter estimate uses the orthogonal crossing shown in tangential view in \textbf{b}. Colour along each intracortical axon is the conduction delay accumulated from that axon's own entry point (key above the sheet), which increases continuously with distance; for H$_1$ the delay to column~A is smaller than the delay to column~B ($d_A < d_B$, brackets). The core of each column is a cluster of local terminals with near-equal delays converging on an L4 stellate cell, which is the synchronous integration site; column boundaries (dashed) run perpendicular to the direction of axonal extension. Below, the same H$_1$ neuron fires at two latencies under different stimuli---earlier for a preferred stimulus ($t_1$), later for a non-preferred one ($t_2$). Because $t_1 + d_B = t_2 + d_A$, the earlier spike arrives in register with other LGN input at the more distal column~B, and the later spike at the more proximal column~A, both inside one coincidence window $\tau$ (shaded). Neighbouring columns therefore read out different firing latencies of the same LGN neuron, and the arrangement of columns is continuous. \textbf{b}, The construction fixes a column's shape as well as its size. A tangential view of L4, looking down on the sheet that \textbf{a} cuts through from the side. Each afferent's iso-arrival-time contours (dashed) run perpendicular to it and are spaced one window's conduction apart, so a target lying between one pair receives that afferent's input inside a single window; the column is where one such strip from each converging afferent overlaps, and is therefore $v \times \tau$ across in every direction an afferent constrains. Two orthogonal afferents -- the pair the estimate adopts -- give a square cell; the two diagonals drawn here clip its corners to an octagon; and afferents arriving at many angles drive it to the inscribed disc (dotted), of the same diameter. The approximately circular cross-section of a column in tangential section is thus a consequence of the same construction that fixes its diameter. \textbf{c}, The diameter that follows, Eq.~\ref{eq:column_diameter}: predicted column diameter against conduction velocity, banded over the effective coincidence window that feed-forward inhibition can impose ($\tau = 1$--2~ms). At $v \approx 0.3$~m/s (dashed line), the speed measured for thalamocortical axons between the white matter and L4, the relation gives 300--600~$\mu$m (bar). Overlaid are the three areas that pair a horizontal velocity attributed to monosynaptic transmission with a column defined by thalamic afferents. V1, monkey: the median of orthodromic single-unit speeds, 0.33~m/s (no spread reported), against single-eye ocular dominance columns 395--670~$\mu$m wide. Barrel, mouse: the mean L4 velocity from voltage responses of parvalbumin interneurons in slices, $142~\mu$m/ms, bar $\pm 1$~SD across slices, against barrels from 124~$\mu$m, an in-slice width and so a lower bound, to about 300~$\mu$m. Cat 17: the 0.3--0.6~m/s of the supragranular pathway, recorded in vivo, against ocular dominance columns 400--500~$\mu$m wide, half their period; no central value is reported, so the two ranges cross at their middles without a marker. Velocity bars show the variation each study reports across its recordings or slices. Sources, and the other areas in which both quantities have been measured, are in \textbf{S7 Text}.}
\label{fig:column}
\end{wfig}

\section{Discussion}

The delay-signature framework holds that physical axonal conduction delays act as a
key matrix for selective sequence decoding, that coincidence detection turns
arrival synchrony into an all-or-none output, and that a single scalar---the
dispersion of the delay set---shifts neurons from detecting events to detecting
sequences. Each of the three claims is anchored to a quantity an anatomist can
measure, and it is that anchoring, more than any individual number, that we
take to be the contribution.
For the third, the anchor is geometry. The shift follows from the geometry of
arrival times, so the calculation needs only the arrival statistics and the
gate, fits nothing, and turns a measured delay distribution into a prediction of
the detectors it produces.

\subsection{State sensitivity as a test of delay compensation}

Sequence detection buys its selectivity with a millisecond timing tolerance,
and that tolerance is measurable in the intact animal. Circuits relying on
wide-dispersion sequence detection should be measurably more temperature- and
state-sensitive than those relying on narrow-dispersion event detection. Near
the codable bound they should also be asymmetrically so. Where a projection's
dispersion is close to the interval it encodes, the detectors it supports
mostly fall short of compensating that interval, so slowing conduction costs
fewer of them than speeding it: 74\% against 46\% survive a 5\% change when the
dispersion equals the interval, and 69\% against 62\% when it exceeds the
interval by a third. Wider projections lose the asymmetry. A slowing advantage
that appears near the bound and vanishes beyond it is a signature of selection
by delay compensation, which makes it a discriminator between this account and any scheme that reads spike timing
without delay compensation.

The simulations above isolate delay scaling; in the intact brain, temperature
and neuromodulatory state would act through additional mechanisms beyond
simple delay changes---ranging from changes in membrane excitability and
synaptic efficacy to the recruitment of homeostatic compensation. The
prediction should therefore be read as an isolated-component estimate.

\subsection{The hard threshold: selectivity, sparsity, and failure under jitter}

The hard calcium threshold is what gives the model a silent, tuned
subpopulation at all: against a soft-gated control on identical delays,
connectivity and stimuli, it raises population selectivity from $0.020$ to
$0.703$ on clean input and still holds $0.375$ at $5$~ms of jitter, while the
soft rule fires for target and distractor alike (\textbf{S3 Fig}). Tolerance to
jitter and the rejection of chance coincidences come from the sparsity of $Q$,
maintained upstream (\textbf{S2 Fig}). The threshold sets how the system fails:
the tuned pool loses its silence to the reversed order and keeps its preference
for the target (\textbf{S3 Fig}).

\subsection{The precision--robustness trade-off of the coincidence window}

The simulations fix the coincidence window at about 1~ms, the narrow end of the windows measured directly.
In the cell, inhibition sets the width, and inhibition moves with recent activity and with neuromodulatory state.
In L4 of barrel cortex the window for thalamic input widens from about 1~ms toward 10~ms as feed-forward inhibition depresses under repetitive stimulation, and the change alters the temporal precision of the cortical response to whisker deflection \citep{gabernet2005somatosensory}; in dentate granule cells, noradrenaline narrows the window to $\sim$4~ms by enhancing CCK$^+$ inhibition \citep{glovaci2026sparsification}.
A circuit can therefore move along a trade-off between precision and robustness.

At the narrow end each input contributes a single precisely timed spike, and the branch fires only for the delay configuration it matches.
Firing is sparse: the same information travels in fewer spikes, and so at lower metabolic cost, consistent with the low firing rates of cortical neurons and with the energy budget that limits them \citep{barth2012experimental,lennie2003cost}.
The price is tolerance: the delay error the code can absorb is of the order of the window itself (Fig.~\ref{fig:limits}f), so a narrow window leaves conduction a narrow budget.
This is the regime the simulations occupy, and it suits encoding, where a new sequence must be told apart from its neighbours.

A wider window accepts more delay configurations as matches for a given input, so keys tuned to neighbouring intervals begin to fire.
In exchange it tolerates the jitter, delay drift and changes in conduction velocity that cost the millisecond window much of its matched population (Fig.~\ref{fig:limits}d, e).
Once several spikes from each input fall within the window, their number matters as well as their timing, and the readout moves toward a rate code.
That exchange may suit memory retrieval or search, where the sequence is already stored and the downstream neuron need only report whether it is present, and where there may be greater tolerance for imprecise input and for the co-activation of neighbouring, possibly irrelevant memory circuits.
No simulation here varies the window. The precision side of the trade-off has been measured, in barrel cortex; the robustness side is argued from the mechanism.

\subsection{The codable bound and the gamma period}

The upper bound of the codable interval is set by $\Delta$, and at the 20--30~ms spread usually attributed to unmyelinated horizontal fibres, this bound falls close to one gamma cycle (30--80~Hz, hence 12--33~ms). This suggests that the gamma period may reflect the delay statistics at the circuit level: spikes separated by more than one cycle fall outside the compensable range, and associating them requires cross-cycle phase locking or theta-nested rhythms. The 20--30~ms spread is inferred from the general range of cortical conduction delays; the nearest direct measurement available comes from rabbit callosal axons \citep{swadlow1985physiological}, most of which conduct at the slow velocities of unmyelinated or thinly myelinated fibres, like the horizontal fibres in question, and so provide a proxy for their delay statistics. The same range is consistent with the corticothalamocortical loop delays discussed in \textbf{S6 Text}.

\subsection{Relation to delay learning and to other temporal readouts}

In this framework conduction delays are physical givens, and learning selects
among them. Delay learning in spiking networks adjusts the delays themselves,
and polychronization tunes weights over a fixed delay graph. The difference is
one of degree, because conduction delays are plastic too: neuronal activity promotes
oligodendrogenesis and adaptive myelination \citep{gibson2014neuronal}, motor
learning requires active central myelination \citep{mckenzie2014motor}, and
vesicle release regulates myelin sheath number on individual axons
\citep{mensch2015synaptic,fields2015new}.
Myelin remodelling large enough to alter conduction most likely takes days to
weeks \citep{almeida2017myelinated}, so delays change more slowly, and
through a different mechanism, than the synapses that select among them.
The developmental prediction sharpens accordingly: if myelination is itself
activity-dependent, the redundant pool is partly re-tuned as well as pruned,
and the timescale separating the two processes is measurable.

Myelin can also be the source of the dispersion itself.
The branch reads only arrival times, so it does not matter whether two delays differ because their axons differ in length, in myelination, or in both.
A projection of similar path lengths can therefore still supply a wide $\Delta$, if its short delays run through myelinated segments and its long ones through unmyelinated segments.
Models of activity-dependent myelination already give it a role in timing: small changes in conduction velocity shift spike arrival enough to alter oscillation frequency and the coupling between oscillators \citep{pajevic2014role}, conduction-delay statistics adjust themselves to stabilize large-scale phase synchrony \citep{noori2020activity}, and the coordination of spike timing that myelin plasticity produces raises firing rates and correlations \citep{talidou2022homeostatic}.
The dispersion result carries that role from timing to computation: plasticity that changes the delay statistics of a projection changes its $\Delta$, and with it the class of temporal feature the projection's targets detect.

The tempotron \citep{gutig2006tempotron} and chronotron \citep{florian2012chronotron} show that a single thresholded unit can learn spike-timing-based decisions, and liquid state machines \citep{maass2002real} obtain temporal selectivity from generic recurrent dynamics. Each learns under supervision: a teacher signal states, pattern by pattern, the output the unit or its readout should have produced, and the error against it drives the weights.
Selection needs no teacher: the rule that makes it is local to the branch, which is potentiated on coincident arrival independently of somatic output.

\subsection{The attention analogy}

The labels $Q$ and $K$ come from attention mechanisms, in which a query is
matched against keys to decide which inputs are integrated, and the
correspondence extends to that logic and no further. Attention-like
computation has also been placed in neuron--astrocyte synapses
\citep{kozachkov2023building} and used to derive hippocampal place and grid
codes \citep{whittington2022relating}. Here the keys are conduction delays
fixed by anatomy, the gate is an all-or-none threshold that makes sparsity
intrinsic, and plasticity selects among existing pathways. One property carries
content beyond the analogy. Attention is permutation-invariant,
so a Transformer sees order only through a positional encoding added to every
token, a choice made once for the whole model. A delay signature reads order
natively and branch by branch: narrow $\Delta$ makes a unit likely to read a set,
wide $\Delta$ a sequence, and a population at intermediate dispersion holds both kinds at once
(Fig.~\ref{fig:dispersion}b). Which units should be order-sensitive, and by how
much, is a question the delay framework poses and the Transformer architecture
has no parameter for. The integrator neuron also admits a predictive-coding
reading, in which the distal match is a contextual prediction that the soma
checks against the content that arrives (\textbf{S2 Text}, \textbf{S3 Text});
none of the results here tests it.

\subsection{Limitations}

The model abstracts away dendritic geometry, ion-channel distribution, and the subtype-specific gating of branches by distinct interneuron populations---PV$^+$ basket cells sharpening the perisomatic window, SOM$^+$ Martinotti cells gating the very distal inputs that undergo matching---so the distal-versus-proximal dichotomy used here is coarser than the real compartmentalization. The transmission efficiency $\eta_j$, the distal gain $\alpha$, the threshold $\theta$ and the plasticity of $K$ are all neuromodulated in vivo and static here. The modelled plateau lasts 15~ms, against 100--200~ms for the biological one \citep{larkum1999new}; lengthening it to 100~ms leaves every class fraction within its confidence interval, because a plateau ends at the first somatic spike (\textbf{S1 Text}). Axonal conduction is treated as the primary source of dispersion in $K$, with the latency from somatic integration to spike held fixed; in local recurrent circuits such as CA3 and L2/3, where axonal segments are short and activity re-enters the same network repeatedly, membrane integration time would contribute a comparable and neuromodulable component to the effective delay, and that dynamic extension of $K$ is not modelled.

Two parameters carry more weight than the measurements behind them. The canonical $\tau_{\mathrm{w}} \approx 1$~ms sits at or below the bottom of every window that has been measured directly. In L4 of barrel cortex, the compartment the column prediction concerns, the window is that narrow only for input on a rested thalamocortical pathway, and repetitive stimulation widens it toward 10~ms \citep{gabernet2005somatosensory}; that measurement was made in slices, and where an awake animal sits within the range is not known. The horizontal velocity is less secure still. Every available estimate is read from postsynaptic responses, and in monkey V1 the latency of orthodromic responses does not grow with distance \citep{zhaoping2025conduction}, so the speed computed from it rises with the distance measured over, from a median of 0.33~m/s in samples mostly within 2~mm \citep{girard2001feedforward} to $1.38 \pm 0.15$~m/s at 3~mm. The relation sets the width of columns where afferents define them and is silent on whether a cortex forms them: rat V1 has monosynaptic horizontal fibres conducting at 0.15--0.55~m/s \citep{murakoshi1993electrophysiological}, within the range of the columnar areas in \textbf{S7 Text}, and no orientation columns \citep{ohki2005functional}.

Two structural claims rest on argument. The first is that somatic integration is conjunctive; a model experiment shows that no weighted sum of non-negative drives can reproduce the ordering a conjunctive soma produces, which shows that a conjunctive soma fits the architecture and leaves open whether real pyramidal neurons implement one (\textbf{S4 Text}). The second is the scaling of the redundant pool: exact pairwise matching over $N$ inputs would demand on the order of $\Delta^{\,N-1}$ candidate branches, which the ``tens to hundreds'' anatomy cannot supply for large $N$. This demand can be partially alleviated through hierarchical stacking and relay neurons (\textbf{S6 Text}).

The model also treats the downstream neuron as quiescent when $Q$ arrives, so that coincidence detection determines whether the soma fires. Cortical neurons are rarely silent. Their inputs fire spontaneously, they carry ongoing excitatory and inhibitory drive and fluctuate subthreshold, and they sit in self-sustaining recurrent circuits. The first of these the model can test. With every input firing as an independent Poisson process, fewer than 5\% of a five-input gate's sequence detectors answer a single event at rates up to 3~Hz, and fewer than 5\% of a three-input gate's up to 0.5~Hz (\textbf{S1 Text}). Of the two gates Fig.~\ref{fig:limits} reports, the strict one, with its tolerance of 0.55~ms, is therefore the one suited to spontaneous rates of a few hertz. In the rest of that regime a distal $Q$ signal would act less as a trigger than as a selector of which downstream neurons participate. Whether the framework survives that change of regime is open, and the question is sharpest for long-range cross-areal projections, which cannot plausibly supply a redundant $K$ pool the size a local circuit offers.

Finally, the framework has not yet been tested against recorded neural data: the test needs two measurements from one projection, the distribution of its conduction delays and the millisecond order tuning of its targets, and we know of no public dataset that pairs them.
Antidromic recordings measure a projection's delays without its targets' tuning \citep{swadlow1985physiological}.
The largest public reconstruction that follows axons to targets with recorded responses spans about a cubic millimetre of mouse visual cortex \citep{microns2025functional}; it truncates the horizontal fibres that would supply a wide dispersion, since 20~ms at 0.3~m/s needs paths differing by 6~mm, and its calcium imaging cannot resolve millisecond order.
Measuring both for one projection is the cheapest route to an empirical anchor, and it is the first item in Box 2.

\begin{pointbox}{Box 2. Testable predictions}
Each claim advanced here carries a specific measurement that would support or
challenge it, and none requires a technique that does not currently exist.

\medskip
\noindent
\textit{i. The dispersion of a projection predicts the temporal tuning of its
targets.} Targets of a narrowly dispersed projection should respond to single
volleys and ignore their order. Targets of a widely dispersed one should include
order-selective neurons, none of which responds to an interval much longer than
the projection's dispersion: that is the codable bound, and the gate barely
moves it. The core claim is challenged if order selectivity appears where
$\Delta$ is narrow, or extends well beyond $\Delta$. The same measurement also reveals whether
$\Delta$ falls in the $\sim$20--30~ms range, in which the codable bound aligns
with the gamma period.

\medskip
\noindent
\textit{ii. Column width predicts the L4 coincidence window.} Where afferents
define a column, its width divided by their conduction velocity inside cortex
gives the window: 1.2--2.0~ms in macaque V1, 0.9--2.1~ms in mouse barrel
cortex and 0.7--1.7~ms in cat area 17. Published velocities cannot test this, because the way $v$ was measured
moves the implied window two- to fourfold (\textbf{S7 Text}); the test needs
$v$ measured directly in afferent arbors in L4, over less than a millimetre, by
one method across several columnar areas. It is challenged if
$d_{\mathrm{column}}/v$ so measured varies across areas while the awake window
of iii does not.

\medskip
\noindent
\textit{iii. In the awake animal, the L4 coincidence window for stimulus-evoked
thalamic input is of order one millisecond.} In slices it ranges from about 1 to
10~ms, widening as the thalamocortical pathway is driven repeatedly
\citep{gabernet2005somatosensory}; the framework requires the narrow end during
sensory coding, and therefore predicts it. It is challenged by an
effective window substantially broader than 2~ms in the awake animal, together
with a failure of manipulations of that window to alter columnar feature
tuning.
\end{pointbox}

\subsection{Conclusion}

The results support a compact claim. Heterogeneous conduction delays are a
physical resource for computation, and the single scalar
that describes their spread shifts a neuron from detecting events to detecting
sequences, bounds how long a sequence it can represent, and sets how much timing
precision it must be given. The scalar is
set by axonal length and myelination, which maps a computational distinction
onto an anatomical one. The precision it demands, of order a millisecond,
matches the window to which feed-forward inhibition holds L4 and the window that
column width divided by conduction velocity implies. The framework is therefore
testable in the delay statistics of real projections and in the coincidence
window of the awake animal, which column anatomy predicts area by area.

\section{Materials and methods}

The code behind every reported result, with the exact command lines and the
result tables, is archived at Zenodo \citep{bi2026code}.
All simulations are discrete-time on a fixed 1~ms step, so delays, time
constants, plateau durations and refractory periods are counts of steps
numerically equal to milliseconds. The exception is the precision sweep, which
perturbs delays by fractions of a millisecond and so runs on a 0.1~ms step.
There, spike offsets and delays are drawn on that grid, every duration is
scaled to keep its length in milliseconds, and the distal gain $\alpha$ is
divided by ten, so that the soma integrates the same charge per millisecond
(\textbf{S1 Text}). Within a step each distal branch sums its
delayed, weighted input onto a local membrane decaying with
$\tau_\mathrm{den} = 2$~ms; if that sum exceeds $\theta$ the branch opens a
plateau of amplitude $\tanh(\mathrm{Ca}_j - \theta)$ and resets its local
membrane. The soma then decays with $\tau_\mathrm{mem}$, adds the proximal
drive together with $\alpha$ times the summed branch currents less any
inhibition, and spikes on reaching $\theta_\mathrm{soma} = 1.0$, after which it
resets to rest, holds a 2~ms refractory period and terminates its own plateaus.
Two parameter sets are used. The two-input single-branch simulations take
$\tau_\mathrm{mem} = 5$~ms, a 10~ms plateau and a constant subthreshold
proximal drive of 0.10 per step. The population simulations take
$\tau_\mathrm{mem} = 10$~ms and a 15~ms plateau, with no proximal drive at all,
so that firing is attributable to distal coincidence alone. Both take
$\alpha = 2.5$ and unit branch-to-soma efficiency.

Connectivity is drawn once per seed: each feature projects to each branch with
probability 0.1, a realized connection carries weight 1.0 and an unrealized one
weight 0, and every connection draws its own delay uniformly from the inclusive
integer range $[0, \Delta]$, so dispersion $\Delta$ is the maximum delay and
$\Delta = 0$ an entirely synchronous projection. The two-input simulations
instead connect both inputs to the single branch, with delays hand-set to
$(5, 0)$ in the worked example and drawn on $[0, 20]$ in the population-noise
sweep. Unit weights make the calcium threshold a count of coincident inputs
directly: $\theta = 1.5$ requires two, $\theta = 2.5$ three and $\theta = 4.5$
five. Because the window is graded rather than a boxcar, being set by
$\tau_\mathrm{den}$, two inputs one step apart deliver $1 + e^{-1/2} = 1.607$
rather than 2; $\theta = 1.8$ therefore also requires two inputs but forbids
that near miss, and is the exact-coincidence operating point of the
population-noise sweep. A two-input branch cannot reach the three- and
five-input thresholds at all.

The two-event stimulus recruits two disjoint sets of 40 neurons from a
200-neuron feature population, each spike offset drawn uniformly from
$[0, 3]$~ms after its volley onset, the first volley at 5~ms (10~ms in the
discriminability comparison) and the second after the inter-event interval, in
a trial of 60~ms (70~ms for the discriminability and precision sweeps, 100~ms
for the aliasing sweep). The single-event controls, the order reversal and
every retiming are re-rendered from one latent description and so carry
identical spike identities and within-volley offsets. The two-input stimulus
places A at 5~ms and B at 10~ms in a 30~ms trial. Jitter displaces each spike
time by an independent draw of $\mathcal{N}(0, \sigma)$ rounded to the nearest
step, $\sigma$ swept from 0 to 5~ms, with spikes leaving the window dropped.
Spurious input places an extra spike in each (step, feature) bin independently
with probability $d$ up to 0.5, superimposed by taking the maximum so that a
spurious spike landing on a target spike leaves one spike there; false
positives are measured on matched trials from the same noise process with the
target removed, which leaves them undefined in the jitter arm. The precision
sweep perturbs delays rather than spikes, scaling each by an independent draw
of $1 + \mathcal{N}(0, s)$ or all by a common factor $k$ and re-rounding to
whole steps.

Populations are 1000 neurons for the discriminability and dispersion sweeps,
2000 for the aliasing and precision sweeps, and pools of 200 for the
population-noise sweep. The $\theta = 4.5$ arms enlarge them to 50{,}000 for
the discriminability comparison, 20{,}000 for the aliasing sweep and
120{,}000 for the precision sweep, and the conjunctive-soma experiment of
\textbf{S4 Text} uses 40{,}000. All simulations run 20 independent seeds, 0
to 19, each seeding the one generator that draws the stimulus, the connectivity
and all noise. The discriminability and population-noise sweeps repeat 60 and
50 noisy trials per seed, the stimulus and connectivity in the former being
drawn once and reused across both gating rules so that the comparison is
paired; the others evaluate the whole population once per seed and take the
population fraction as that seed's observation.

Neurons are selected functionally, never by connectivity. The dispersion,
aliasing and precision sweeps keep those firing for the two events together and
for neither alone, and the dispersion sweep then measures order selectivity as
the further fraction of those also silent to the reversed order, which takes no
part in the selection. The discriminability comparison keeps those firing for
the clean target and not the clean distractor, selects them once under hard
gating and scores both gating rules on the same neurons. The population-noise
sweep screens for neurons firing to the clean two-spike target and uses the
unmatched remainder as its false-positive baseline.

The class fractions of the dispersion sweep were also computed without
simulation. Each of an event's 40 features connects to the branch with
probability $p$; a connected spike leaves at its volley's onset plus an offset
uniform on $\{0, \dots, 3\}$ and arrives after a delay uniform on
$\{0, \dots, \Delta\}$, and a feature arrives at most once. The branch
potential obeys $x_t = e^{-1/\tau_\mathrm{den}} x_{t-1} + N_t$, with $N_t$ the
number of arrivals in step $t$, and the neuron fires if $x_t$ exceeds
$\theta + \epsilon$. The margin $\epsilon = 0.049$ is how far the potential
must pass threshold for the resulting plateau, of amplitude
$\tanh(x_t - \theta)$, to bring the soma to its own threshold before the
plateau ends. The probability of never firing follows exactly, by propagating
step by step the joint distribution of the potential, on a grid of $10^{-3}$,
and of the number of each event's features yet to arrive. Firing to event 1
alone ($F_1$) and to event 2 alone ($F_2$) involve disjoint features and are
independent. Below threshold the two potentials add, so silence to the pair
implies silence to each. The class fractions are then
$P(\mathrm{sequence}) = P(\bar F_1) P(\bar F_2) - P(\bar F_{12})$,
$P(\mathrm{event~1}) = P(F_1) P(\bar F_2)$,
$P(\mathrm{dual}) = P(F_1) P(F_2)$ and
$P(\mathrm{silent}) = P(\bar F_{12})$, where $F_{12}$ is firing to the two
together. Every parameter is the simulation's own, and none is fitted.

Each seed contributes one observation to each reported quantity, whose value is
the mean over the 20 seeds and whose interval is a 95\% percentile bootstrap of
that mean from 10{,}000 resamples; intervals are therefore over seeds and not
over trials or neurons, and percentile rather than normal-theory because
several quantities are proportions near zero. The crossover and peak dispersions
are read off the seed-pooled curves rather than seed by seed, since a single
population at $\theta = 4.5$ rarely holds a sequence detector near the
crossing. Their intervals, and those of the slopes fitted to them, come from
4{,}000 bootstrap resamples of the seeds. Order-selectivity counts are pooled
over seeds rather than averaged per seed, the counts one population yields at
$\theta = 4.5$ being too small to ratio singly. Full
per-experiment parameterization is given in \textbf{S1 Text}.

\subsection*{Use of generative AI}
The authors used Claude (Anthropic) to verify the ideas in this work and to
improve the manuscript's language.
To verify the ideas, it wrote the code in the companion package under the
authors' direction: the simulator, the exact calculation from the arrival
statistics shown in Fig.~\ref{fig:dispersion}, the experiment and figure
scripts, and the test suite.
It also ran the simulations, checked the values quoted in the text and captions
against their output, searched the literature, and checked citations against
Crossref and PubMed.
For the manuscript, it corrected grammar and drafted and edited parts of the
text.
Every figure is drawn by these scripts, and none contains AI-generated imagery.
The framework, the model and the interpretation of the results are the authors'
own.
The authors directed and reviewed all AI-assisted work, and a test suite checks
the simulator against an independent single-neuron reference implementation.
The authors take full responsibility for the content of this publication.

\section*{Supporting information}

\paragraph*{S1 Fig.}
{\bf Redundancy-dependent plasticity of the delay signature.} The same upstream
sequence reaches many downstream branches through many delay configurations;
only the branch whose delays compensate the spike-time offsets receives
synchronous arrival and a calcium plateau, and only its connections are
retained. Arrival-time constructions for a matched and an unmatched branch are
drawn on a common time scale.

\paragraph*{S2 Fig.}
{\bf Population coding and $Q$-sparsity dependence.} Detection rate and
false-positive rate against Poisson noise density on the $Q$ signal, for pools
of 200 integrator neurons with delays drawn uniformly on $[0, 20]$~ms, together
with the analytic account of the match rate.

\paragraph*{S3 Fig.}
{\bf Delay compensation discriminates temporal order, and hard gating buys
sparsity.} A two-input worked example against a leaky integrate-and-fire
control with the same membrane parameters and no delays, whose trajectories
superimpose across the two orders; then hard against soft gating on identical
delays, connectivity and stimuli. Hard gating is thirty-fivefold more selective
at the population level, emits a fraction of the downstream spikes the soft rule
does, and is the only one of the two with a silent baseline; it loses roughly
half that selectivity under 5~ms of jitter.

\paragraph*{S1 Text.}
{\bf Detailed simulation parameters.} Full parameterization of every simulation
reported in the Results, one subsection each: population and feature-pool
sizes, connection probability, stimulus construction, neuron parameters, the
swept grids, and the enlarged populations of the $\theta = 4.5$ arms, together
with the script that produces each. Includes the aliasing and precision results
at the strict five-input gate, the vote-gain robustness analysis and the
leaky integrate-and-fire control parameters.

\paragraph*{S2 Text.}
{\bf Predictive coding in a dendritic implementation.} Systematic comparison
with classical predictive coding, mechanisms of error signal conversion,
cross-regional error propagation, error-driven memory circuit migration, and
the correspondence with the free-energy principle.

\paragraph*{S3 Text.}
{\bf Index--content separation and hierarchical Q-cycles in cortical and
hippocampal circuits.} A mechanistic
demonstration of the index-library principle using a reduced CA3 network model,
followed by concrete biological instances of
hierarchical Q-cycles in hippocampal circuits (DG$\rightarrow$CA3$\rightarrow$CA1),
recurrent microcircuits, and LGN temporal formatting.

\paragraph*{S4 Text.}
{\bf Somatic integration is conjunctive.} A model experiment.
Any somatic rule that sums non-negative drives is forbidden from firing less often
to mismatched content than to absent content; a conjunctive soma does exactly
that, reaching a veto index of 0.763 while holding matched firing at 100\%,
whereas a weighted sum never exceeds 0.000 at any inhibition strength. The
result shows that a conjunctive soma fits the architecture and that no additive
rule can; whether real pyramidal neurons implement one is open.

\paragraph*{S5 Text.}
{\bf Comparison with related frameworks.} A systematic comparison across five
dimensions---temporal selectivity, gating mechanism, structural plasticity,
index--content separation, and biological substrate---against delay-line,
dendritic, attention-based and spiking-network accounts.

\paragraph*{S6 Text.}
{\bf Systematic delay-signature generation through relay divergence for variable $Q$
signals.} How a circuit can cover the positions a feature may occupy in a
sequence when the timing of that feature is not fixed in advance: project it
through parallel relays, so that multiple delayed copies meet different $K$
compensations downstream. The cerebellar granular layer is discussed as a
candidate substrate. Not simulated.

\paragraph*{S7 Text.}
{\bf Column diameter: evidence, uncertainties and corollaries.} Every
cortical area in which a horizontal conduction velocity and a column width have
both been measured, with the coincidence window each pair implies; the
propagation of the uncertainties in $v$, in $\tau_{\mathrm{w}}$ and in the
crossing geometry, which widens the prediction to 100--1000~$\mu$m; and two
corollaries of the same construction that cost no further assumption---that
column boundaries should run perpendicular to the horizontal axonal axis, and
that a column's approximately circular cross-section follows from the same
geometry that fixes its diameter.

\ifarchivebackmatter

\section*{Data availability}
All code needed to reproduce every reported number is openly available at
\url{https://github.com/bicheng2028/Transformer_attention_neuron_dendrites}
and archived at Zenodo as release v1.0.0 \citep{bi2026code}.
The companion package sits under \texttt{delaykey-main/}: the
integrator-neuron model in \texttt{delaykey/}, one script per experiment under
\texttt{experiments/}, and one script per figure under \texttt{figures/}. Each simulation reported in the Results, and the
conjunctive-integration experiment of \textbf{S4 Text}, has a script, and the
package README carries the table mapping manuscript experiments onto script
names, which do not coincide.
The CA3 index-library demonstration of \textbf{S3 Text} runs in the notebook
\texttt{quickstart\_experiments\_partial.ipynb} at the repository root and
uses PyTorch. The two-input worked example of
\textbf{S3 Fig}a--c---one branch, two inputs, delays hand-set to
$(5, 0)$---is drawn by its own generator from illustrative traces rather than
by an experiment script. The claim it makes is nonetheless asserted against the
production simulator: a named test in the suite runs the
same configuration through the population simulator and checks all three of its
stated properties. Each experiment writes a tidy CSV carrying its own seed, and
each figure is regenerated from those CSVs alone, so any single reported value
can be traced to the run that produced it and reproduced in isolation. Because
the scripts' command-line defaults do not reproduce the reported grids, a
\texttt{run\_all.sh} in the package records every exact invocation behind every
reported CSV, including the threshold-consistency arms at $\theta = 4.5$. The
test suite verifies the vectorized simulator against a transparent
single-neuron reference implementation across both gating rules and both
proximal-drive paths. No data other than simulation output underlie the
findings.

\section*{Funding}
The authors received no specific funding for this work.

\section*{Competing interests}
The authors declare no competing financial interests or personal relationships
that could have appeared to influence the work reported in this paper.

\section*{Author contributions}
\textbf{Cheng Bi:} Conceptualization, Methodology, Software, Formal analysis,
Investigation, Visualization, Writing -- original draft, Writing -- review \&
editing. \textbf{Jipeng Sun:} Software, Validation, Formal analysis,
Investigation, Data curation, Visualization, Writing -- review \& editing,
Project administration.

\section*{Acknowledgments}
The authors thank colleagues for helpful discussions.

\fi

\bibliographystyle{elsarticle-harv}
\bibliography{references}

\end{document}